\documentclass[a4paper,12pt]{article}
\usepackage{amssymb}
\usepackage{amsmath}
\date{}
\title{Schwinger Effect in Non-parallel D1-branes:\\
A Path Integral Approach}
\author{A. Jahan and D. Kamani\\Faculity of Physics, Amirkabir University of Technology\\P. O. Box: 15875-4413, Tehran, Iran\\jahan,\,kamani@aut.ac.ir}
\begin{document}
\maketitle
\begin{abstract}
We study the Schwinger effect in a system of non-parallel D1-branes for the bosonic strings using the path integral formalism. We drive the string pair creation rate by calculating the one loop vacuum amplitude of the setup in presence of the background electric filed defined along one of the D1-branes. We find an angle dependent minimum value for the background field and show that the decaying of vacuum into string pairs takes place for the field above this value. It is shown that in $\theta\rightarrow\frac{\pi}{2}$ limit the vacuum becomes stable and thus no pair creation occurs.
\end{abstract}
\section*{\large 1\quad Introduction}
The dynamics of string theory in a background gauge field has been extensively studied for a long time [1-3]. Such a considerations are supported by the fact that the open string spectrum includes the massless gauge fields carrying the corresponding charges at its boundaries [4]. In particular it has been revealed that the string theory vacuum signals instabilities for a string interacting with a constant background electric field. Infact this is the stringy counter part of the well-known Schwinger effect in QED [5-15]. In more recent studies the emphasis is put on the role of the lower dimensional D-branes on which the open strings end. It seems that string pair creation takes place in the D$p\,$-D$p$ and D$p\,$-$\overline{\textrm{D}}p$ systems in type-IIA or type-IIB setup, although the criterion for occurrence of the vacuum decay plays more drastic role in the D$p\,$-$\overline{\textrm{D}}p$ system due to some exponential factors depending on the background configurations [14]. It is also pointed out that a fixed magnetic field can greatly enhance the pair creation rate in the case of a weak electric background [15].\\\
In previous work the path integral formalism was engaged to derive the one loop vacuum amplitude of an angled D1-branes system in the bosonic sting setup by one of the authors [16]. Here we propose to study the Schwinger effect in such a system by calculating the zero point energy using the similar technique presented in [16]. As is expected, the pair creation rate depends on the angle between the D1-branes,  $\theta$. A more interesting feature of the model is that there is an angle dependent minimum value for the electric field and for the field below this value the string pair creation disappears. The vacuum becomes stable in $\theta\rightarrow\frac{\pi}{2}$ limit implying that no string pair creation is expected to occur. \\
Throughout this work we assume the Euclidean signature for both of the world-sheet and space-time manifolds, but immediate continuation to the Lorentzian signature for space-time coordinates is assumed after integration over the bosonic degrees of freedom.
\section*{\large 2\quad One Loop Vacuum Energy}
We begin with the bosonic string action in the $d$ dimensinal space-time by writing it as
\begin{equation}\label{1}
S=\frac{T}{2}\int d^2\sigma \partial_aX^\mu\partial^aX_\mu+S_{ghost}[b,c].
\end{equation}
At one loop level the annulus and m\"{o}bius diagrams are the only diagrams which contribute to the partition function of open string. Taking into account the contribution made by the annulus diagram (which is the most relevant to the pair creation) the one loop vacuum energy (zero point energy) could be written as [11-13]
{\setlength\arraycolsep{2pt}
\begin{eqnarray}\label{2}
\mathcal F&=&\int_0^\infty \frac{ds}{s}\int^\prime DX^\mu DbDc \,e^{-S_0[X]-S_{ghost}[b,c]},
\end{eqnarray}}
where $X^\mu(\sigma,\tau)=X^\mu(\sigma,\tau+s)$. The prime over the second integral means that the contribution of zero modes are excluded in evaluating the above path integral. In presence of a constant electric background the free energy acquires an imaginary part and the vacuum begins to decay into the string pairs with decay rate given by $w=-2\textrm{Im}\mathcal F$.\\
Now, let us consider a bosonic string stretched between two parallel D1-branes located at a relative distance $Y$. There are 2 degrees of freedom satisfying the Neumann-Neumann (NN) boundary condition $\partial_{\sigma}X_N|_{\partial\Sigma}=0$ and $d-2$ degrees of freedom satisfying the Dirichlet-Dirichlet (DD) boundary condition $\delta X_D|_{\partial\Sigma}=0$. So, the partition function becomes $\mathcal Z=Z_{N}^2Z^{d-2}_{D}Z_{gh}$ where the partition function of ghost fields is given by $Z_{gh}=\frac{T}{2s}Z^{-2}_{N}$. For the ends of string we suppose
{\setlength\arraycolsep{2pt}
\begin{eqnarray}\label{3-4}
X^i(0,\tau)&=&0,\qquad i=2,\cdots,d\\
X^i(\pi,\tau)&=&l_i.
\end{eqnarray}}
Thus for the typical fluctuations $X_{N}$ and $X_{D}$ we write
{\setlength\arraycolsep{2pt}
\begin{eqnarray}\label{5-6}
X_{N}&=&\sum_{m\in\mathbb Z}\sum_{n=0}^\infty\chi_{mn}u_{mn},\\
X_{D}&=&\sum_{m\in\mathbb Z}\sum_{n=1}^\infty\xi_{mn}v_{mn}+\frac{l}{\pi}\sigma.
\end{eqnarray}}
Here, the eigen-modes are $u_{mn}=e^{i\omega_m\tau}\cos n\sigma$ and $v_{mn}=e^{i\omega_m\tau}\sin n\sigma$. Furthermore, we define $\omega_m=m\omega $, where $\omega=\frac{2\pi}{s}$. Therefore, we find for the corresponding actions $S_{N}$ and $S_{D}$ as
{\setlength\arraycolsep{2pt}
\begin{eqnarray}\label{7-8}
S_{N}&=&2\pi\sum_{m=1}^{\infty}\textbf{x}^{\dagger}_{m}\textbf{M}_m\textbf{x}_{m}
+\pi\textbf{x}^{\scriptsize\textrm{t}}_{0}\textbf{M}_0\textbf{x}_{0},\\
S_{D}&=&2\pi\sum_{m=1}^{\infty}\textbf{y}^{\dagger}_{m}\textbf{N}_m\textbf{y}_{m}
+\pi\textbf{y}^{\scriptsize\textrm t}_{0}\textbf{N}_0\textbf{y}_{0}+\frac{1}{2\pi}s\alpha l^2,
\end{eqnarray}}
where $\textbf{x}^{\dagger}_m=(\bar{\chi}_{m0},\bar{\chi}_{m1},...)$, $\textbf{y}^{\dagger}_m=(\bar{\xi}_{m1},\bar{\xi}_{m2},...)$ and $\lambda_{mn}=\omega_m^2+n^2$. The matrix $\textbf{M}_m$ is defined by its elements as
\begin{equation}\label{9}
[\textbf{M}_m]_{nn^\prime}=\frac{1}{4}sT
\left\{\begin{array}{ll}
2\lambda_{m0}\,\qquad n=n^\prime=0\\
\lambda_{mn}\delta_{nn^\prime}\quad n,n^\prime\neq 0.
\end{array} \right.\qquad
\end{equation}
Furthermore, there are $[\textbf{M}_0]_{nn^\prime}=\frac{1}{4}sT\lambda_{0n}\delta_{nn^\prime}$ and $[\textbf{N}_m]_{nn^\prime}=\frac{1}{4}sT\lambda_{mn}\delta_{nn^\prime}$. Integration over the fluctuation $X_{N}$ leads to
{\setlength\arraycolsep{2pt}
\begin{eqnarray}\label{10}
Z_{N}(s)&=&\int^\prime DX_{N}e^{-S_0[X_{N}]}=\prod_{m=1}^{\infty}\int^\prime d\textbf{x}^{\dagger}_md\textbf{x}_md\textbf{x}_0 e^{-S_{N}[\textbf{x}^{\dagger}_m,\textbf{x}_m,\textbf{x}_0]}\\\nonumber
&=&\frac{1}{\sqrt {\textrm{det}\textbf{M}_0}}\prod_{m=1}\frac{1}{\textrm{det}\textbf{M}_m}\\\nonumber
&=&\sqrt{\frac{T}{2s}}q^{-\frac{1}{24}}\prod_{n=1}^\infty\frac{1}{1-q^{n}}.\nonumber
\end{eqnarray}}
where $q=e^{-s}$. In the same way, we obtain
{\setlength\arraycolsep{2pt}
\begin{eqnarray}\label{11}
Z_{D}(s)&=&\int DX_{D} e^{-S_0[X_{D}]}=\prod_{m=1}^{\infty}\int d\textbf{y}^{\dagger}_md\textbf{y}_md\textbf{y}_0 e^{-S_{D}[\textbf{y}^{\dagger}_m,\textbf{y}_m,\textbf{y}_0]}\\\nonumber
&=&\frac{e^{-\frac{1}{2\pi}T l^2}}{\sqrt {\textrm{det}\textbf{N}_0}}\prod_{m=1}\frac{1}{\textrm{det}\textbf{N}_m}\\\nonumber
&=&q^{\frac{1}{2\pi}T l^2-\frac{1}{24}}\prod^\infty_{n=1}\frac{1}{1-q^{n}}.
\end{eqnarray}}
For the distance between D1-branes we have $Y^2=\sum_{i=2}^dl^2_i$. Now, we consider the case of angled D1-branes. In this case the two degrees of freedom, one satisfying NN and the other satisfying DD boundary condition, turn to satisfy the mixed boundary condition. So, the partition function becomes $\mathcal Z=Z_{N}Z^{d-3}_{D}Z_{mix}Z_{gh}$. Hence, with deflection angle $0\leq\theta\leq\pi$, equations (3) and (4) for the ends of string modifies to
{\setlength\arraycolsep{2pt}
\begin{eqnarray}\label{12}
X^i(0,\tau)&=&0,\quad\quad i=2,...,d\\
X^{i^{\prime}}(\pi,\tau)&=&l_{i^{\prime}}\quad\quad i^{\prime}=3,...,d
\end{eqnarray}}
The conditions satisfied by the ends of an open string at the boundaries, imposed by the classical equations of motion, read [16-19]
{\setlength\arraycolsep{2pt}
\begin{eqnarray}\label{15}
\partial_\sigma X^1(0,\tau)&=&0,\\
\partial_\tau X^2(0,\tau)&=&0,\\
\partial_\sigma X^1(\pi,\tau)\cos\theta&=&-\partial_\sigma X^2(\pi,\tau)\sin\theta,\\
\partial_\tau X^2(\pi,\tau)\cos\theta&=&\partial_\tau X^1(\pi,\tau)\sin\theta.
\end{eqnarray}}
This system is T-dual to a magnetized parallel D2-branes [20]. To see this let us T-dualize the setup along the $X^2$ direction upon interchanging $\partial_\tau X^2\leftrightarrow\partial_\sigma X^2$ in (15), (16) and (17). One obtains
{\setlength\arraycolsep{2pt}
\begin{eqnarray}\label{15}
\partial_\sigma X^1(0,\tau)&=&0,\\
\partial_\sigma X^2(0,\tau)&=&0,\\
\partial_\sigma X^1(\pi,\tau)\cos\theta&=&-\partial_\tau X^2(\pi,\tau)\sin\theta,\\
\partial_\sigma X^2(\pi,\tau)\cos\theta&=&\partial_\tau X^1(\pi,\tau)\sin\theta.
\end{eqnarray}}
Hence the set of equations (18-21) characterizes a D2-D2 brane system with magnetic field given by $B=\tan\theta$. For the fluctuations which satisfy the mixed boundary condition of equations (14-17)
we choose the eigen-modes to be $u^a_{mn}=\cos\sigma n_a e^{i\omega_m\tau}$ and $v^a_{mn}=\sin\sigma n_a e^{i\omega_m\tau}$ and expand them as [16]
\begin{equation}\label{18}
\left\{ \begin{array}{ll}
X^1\\
X^2
\end{array} \right\}=\sum_{m,n\in\mathbb Z}\frac{\chi_{mn}}{\sqrt 2}
\left\{ \begin{array}{ll}
u^a_{mn}\\
v^a_{mn}
\end{array} \right\}.
\end{equation}
with common eigen-value $\lambda^a_{mn}=n^2_a+\omega^2_m=(n+a)^2+\omega^2_m$. The number $a=\frac{\theta}{\pi}$ takes the values $0\leq a\leq1$. So, by following the same steps which led to the equations (7) and (8) and by noting that $\lambda^a_{m,-n}=\lambda^{-a}_{mn}$, we get
{\setlength\arraycolsep{2pt}
\begin{eqnarray}\label{19}
S_{mix}&=&2\pi\sum_{m=1}^{\infty}(\textbf{x}^{\dagger }_{+,m}\textbf{M}^a_m\textbf{x}_{+,m}+\textbf{x}^{\dagger }_{-,m}\textbf{M}^{-a}_m\textbf{x}_{-,m})
+\pi(\textbf{x}^{\scriptsize\textrm t}_{+,0}\textbf{M}^a_0\textbf{x}_{+,0}+\textbf{x}^{\scriptsize\textrm t}_{-,0}\textbf{M}^{-a}_0\textbf{x}_{-,0}),
\end{eqnarray}}
where $[\textbf{M}^a_m]_{n n^\prime}=\frac{1}{4}sT\lambda^a_{nm}\delta_{nn^\prime}$. Therefore, we find the partition function as
{\setlength\arraycolsep{2pt}
\begin{eqnarray}\label{20}
Z_{mix}(s)&=&\int^\prime DX^1DX^2 e^{-S[X^1,X^2]}
=\frac{1}{\sqrt{\det\textbf{M}^a_0\det\textbf{M}^{-a}_0}}\prod_{m=1}^{\infty}\frac{1}{\det\textbf{M}^a_m\det\textbf{M}^{-a}_m}\\\nonumber
&=&\frac{q^{\frac{a}{2}(1-a)-\frac{2}{24}}}{1-q^a}\prod_{n=1}^\infty\frac{1}{1-q^{n-a}}\frac{1}{1-q^{n+a}}.\nonumber
\end{eqnarray}}
This leads to the one loop vacuum energy as [16-19]
\begin{equation}\label{21}
\mathcal {F}=\int_0^\infty\frac{ds}{s}\sqrt{\frac{T}{2s}}\frac{q^{\frac{1}{2\pi}T Y^2-\frac{d-2}{24}-\frac{a}{2}(a-1)}}{1-q^a}
\prod_{n=1}^\infty(1-q^{n})^{-d+4}(1-q^{n+a})^{-1}(1-q^{n-a})^{-1}
\end{equation}
In analogy with this expression one obtains the free energy of a magnetized D2-D2 configuration in T-dual picture as
\begin{equation}\label{21}
\mathcal {F}=\frac{BT}{4\pi}\int_0^\infty\frac{ds}{s}\bigg(\frac{T}{2s}\bigg)^{\frac{3}{2}}\frac{q^{\frac{1}{2\pi}T Y^2-\frac{d-2}{24}-\frac{b}{2}(b-1)}}{1-q^b}
\prod_{n=1}^\infty(1-q^{n})^{-d+4}(1-q^{n+b})^{-1}(1-q^{n-b})^{-1}
\end{equation}
where $b=\frac{1}{\pi}\tan^{-1}B$. One must note that switching on the electric flux in a magnetized D2-D2 or (D2-$\overline {\textrm D}$2) system leads to a moving intersecting D1-branes at angle (a moving scissor) in T-dual picture [14, 20-22].
\section*{\large 3\quad  Vacuum Energy: Interaction With Constant Background}
The action of an open bosonic string with charge $q$ located at one of its endpoints (e.g. $\sigma=0$) interacting with an external $U(1)$ gauge field $A_a$ is [1, 2]
\begin{equation}\label{22}
S=\frac{T}{2}\int_\Sigma{d^2\sigma}\partial_{a}X^{\mu}\partial^{a}X_{\mu}
+q\int_{\sigma=0}{d\tau}A_a\partial_\tau{X^a}+S_{ghost}[b,c].
\end{equation}
For a constat electric field with substituting $qE\rightarrow E$, the above expression can be recast in
\begin{equation}\label{23}
S_{int}=\frac{1}{2}qF_{ab}\int_0^s{d\tau}X^a\partial_\tau{X^b}
=-\frac{i}{2}E\epsilon_{ab}\int_0^s{d\tau}X^a\partial_\tau X^b,
\end{equation}
which indicates a mixing between the coordinates $X^0$ and $X^1$. Here, we introduce the action $S_E=S_0[X^0]+S_0[X^1]+S_0[X^2]+S_{int}$, which can be put in another form
\begin{equation}\label{24}
S_E=\frac{T}{2}\int d^{\,2}\sigma Y^{\,\scriptsize\textrm{t}}\mathfrak D Y,
\end{equation}
where $Y=(X^0,X^1,X^2)$ and
\begin{equation}\label{25}
\mathfrak D=
\left(\begin{array}{ccc}
\Box&-i\frac{E}{T}\delta(\sigma)\partial_\tau&0\\
i\frac{E}{T}\delta(\sigma)\partial_\tau&\Box&0\\
0&0&\Box
\end{array}\right).
\end{equation}
In particular for $S_{int}$ we find
\begin{equation}\label{26}
S_{int}=\pi E\sqrt 2\sum_{n=0}\sum_{n^\prime\in\mathbb Z}\sum_{m\in\mathbb Z}m\bar \chi^0_{mn}\chi^1_{mn^\prime}.
\end{equation}
To make the notation more compact, we introduce the matrices $\textbf{I}_m$ and $\textbf{J}_m$ to rewrite the interaction action as
\begin{equation}\label{27}
S_{int}=2\pi\sum_{m=1}^\infty(\textbf x_m^{0\dagger}\textbf{I}_m\textbf x_{m+}+\textbf x_m^{0\dagger}\textbf{J}_m\textbf x_{m-}-\textbf x_{m+}^{\dagger}\textbf{I}_m\textbf x^0_m-\textbf x_{m-}^{\dagger}\textbf{J}^{\scriptsize\textrm{t}}_m\textbf x^0_m).
\end{equation}
where the matrices $\textbf I_m$ and $\textbf J_m$ are defined as
\begin{equation}\label{30}
\textbf I_m=\frac{E_m}{\sqrt 2}\left(\begin{array}{cccc}
1&1 &\ldots\\
1 &1&\ldots\\
\vdots&\vdots&\ldots
\end{array}\right),\qquad
\textbf J_m=\frac{E_m}{\sqrt 2}\left(\begin{array}{cccc}
0&1 &\ldots\\
0 &1&\ldots\\
\vdots&\vdots&\ldots
\end{array}\right).
\end{equation}
and $E_m=mE$. In a similar way, introducing the matrix $\mathcal M_m$ leads to a compact form for the action $S_E$
{\setlength\arraycolsep{2pt}
\begin{eqnarray}\label{28}
S_E&=&\pi\big(\textbf{x}^{\scriptsize\textrm t}_{0+}\textbf{M}^a_0\textbf{x}_{0+}+\textbf{x}^{\scriptsize\textrm t}_{0-}\textbf{M}^{-a}_0\textbf{x}_{0-}+\textbf{x}^{\,0\scriptsize\textrm t}_{0}\textbf{M}_0\textbf{x}^0_{0}\big)
+\sum_{m=1}^{\infty}\zeta^{\,\dagger}_m\mathcal M_m\zeta_m.
\end{eqnarray}}
with $\zeta^{\,\dagger}_m=(\textbf{x}^{\dagger}_{m+},\textbf{x}^{0\,\dagger}_{m},\textbf{x}^{\scriptsize\dagger}_{m-})$ and
\begin{equation}\label{29}
\mathcal{M}_m=2\pi
\left(\begin{array}{ccc}
\textbf{M}^{a}_m&-\textbf{I}_m&0\\
\textbf{I}_m&\textbf{M}_m&\textbf{J}_m\\
0&-\textbf{J}^{\scriptsize\textrm t}_m&\textbf{M}^{-a}_m
\end{array}\right),
\end{equation}
Adding all these together, one finds the partition function as
\begin{equation}\label{31}
Z_E=\frac{1}{\sqrt{\det\textbf{M}^a_0\det\textbf{M}^{-a}_0\det\textbf{M}_0}}\prod_{m=1}^{\infty}\frac{(2\pi)^3}{\det\mathcal M_m}.
\end{equation}
The determinant of matrix $\mathcal M_m$ is calculated in appendix. So, we skip the details and go on by reminding that
\begin{equation}\label{32}
\sum_{n=-\infty}^{\infty}\frac{1}{\lambda^a_{mn}}=\frac{\pi}{2\omega_m}\big[\coth\pi(\omega_m+ia)+\coth\pi(\omega_m-ia)\big].
\end{equation}
This leads us to the final form of the partition function
{\setlength\arraycolsep{2pt}
\begin{eqnarray}\label{33}
Z_E&=&Z_NZ_{mix}\sqrt {1-\frac{E^2}{T^2}}\prod_{m=1}^\infty\frac{(1-e^{-2\pi(\omega_m+ia)})(1-e^{-2\pi(\omega_m-ia)})}
{(1-e^{-2\pi(\omega_m+\beta)})(1-e^{-2\pi(\omega_m-\beta)})},\\\nonumber
&=&Z_NZ_{mix}\sqrt {1-\frac{E^2}{T^2}}\prod_{m=1}^\infty\frac{f_m(\nu|\tau)}{f_m(\nu^\prime|\tau)}.
\end{eqnarray}
where $\nu=a$, $\nu^\prime=i\beta$ and $\tau=i\omega$. For the parameter $\beta$ we have
\begin{equation}\label{18}
\beta=
\left\{ \begin{array}{ll}
i\gamma,\qquad |\alpha|\leq1\\
\epsilon,\qquad\quad 1<\alpha
\end{array} \right.
\end{equation}
where $\epsilon=\frac{1}{2\pi}\cosh^{-1}\alpha$ and $\gamma=\frac{1}{2\pi}\cos^{-1}\alpha$. The parameter $\alpha$ is defined as
\begin{equation}\label{35}
\alpha=\frac{\cos 2\theta+\frac{E^2}{T^2}}{1-\frac{E^2}{T^2}}.
\end{equation}
for which we have $-1\leq\alpha<\infty$. Let's first consider the case $1<\alpha$. The modular property of the Jacobi function ($z=e^{2\pi i\nu}$)
{\setlength\arraycolsep{2pt}
\begin{eqnarray}\label{36}
\Theta_1(\nu|\tau)&=&2e^{\frac{i\pi\tau}{4}}\sin\pi\nu\prod_{m=1}^{\infty}(1-e^{2\pi i\tau m})(1-ze^{2\pi i\tau m})(1-z^{-1}e^{2\pi i\tau m})\\\nonumber
&=&-\frac{e^{-i\frac{\nu^2}{\tau}}}{\sqrt{-i\tau}}\Theta_1\Big(\frac{\nu}{\tau}\Big{|}-\frac{1}{\tau}\Big),
\end{eqnarray}}
and the Dedekind eta function $\eta(\tau)=e^{i\frac{\pi \tau}{12}}\prod_{m=1}^{\infty}(1-e^{2\pi i\tau m})=\frac{\eta(-\frac{1}{\tau})}{\sqrt{-i\tau}}$
provides an equivalent representation for the expression (38) as
\begin{equation}\label{37}
\frac{f_m(\nu|\tau)}{f_m(\nu^\prime|\tau)}=\frac{\sin\pi\nu^{\prime}}{\sin\pi\nu}
\frac{\sin(\frac{\pi\nu}{\tau})}{\sin(\frac{\pi\nu^{\prime}}{\tau})}\frac{f_m(\frac{\nu}{\tau}|-\frac{1}{\tau})}
{f_m(\frac{\nu^\prime}{\tau}|-\frac{1}{\tau})}e^{-\frac{i\pi}{\tau}(\nu^2-\nu^{\prime2})},
\end{equation}
We find
\begin{equation}\label{38}
\frac{\sin(\frac{\pi\nu}{\tau})}{\sin\pi\nu}e^{-\frac{i\pi}{\tau}\nu^2}\prod_mf_m\Big(\frac{\nu}{\tau}
\Big{|}-\frac{1}{\tau}\Big)=\big({2i\sin\theta}\,q^{\frac{1}{12}}Z_{mix}\big)^{-1},
\end{equation}
and since $\mathcal Z=Z_{E}Z^{d-3}_{D}Z_{gh}$, we get
\begin{equation}\label{39}
\mathcal Z=R_{\epsilon}\sqrt\frac{T}{2s}\frac{q^{\frac{1}{2\pi}TY^2-\frac{d-2}{24}+\frac{1}{2}\epsilon^2}}{\sin(\frac{s\epsilon}{2})}
\prod_{n=1}^\infty(1-q^{n})^{-d+4}(1-q^{n+i\epsilon})^{-1}(1-q^{n-i\epsilon})^{-1},
\end{equation}
where the factor $R_\epsilon$ is found to be
\begin{equation}\label{40}
R_\epsilon=\frac{1}{2\sin\theta}\sqrt{\frac{E^2}{T^2}-\sin^2\theta}.
\end{equation}
Beginning from (44) one can easily recover its zero field limit, Eq.(25), by noting that in the limit $E\rightarrow 0$ we have $R_\epsilon=\frac{i}{2}$ and $\epsilon=ia$. The equation (44) has singularities at $s_n=\frac{2\pi n}{\epsilon}$. These singularities lead to an imaginary part for (44), which can be captured with the aid of the well-known formula
\begin{equation}\label{41}
\frac{1}{x-i\varepsilon}=P\frac{1}{x}+i\pi\delta(x).
\end{equation}
So, by noting that $\sin\phi\sim(-1)^n(\phi-\pi n)$ we find the string pair creation rate as
\begin{equation}\label{42}
w(\theta)=R_\epsilon\sqrt{\frac{\epsilon}{\pi}T} \sum_{n=1}^\infty(-1)^{n+1}n^{-\frac{3}{2}}e^{-n(\frac{1}{\epsilon}TY^2-\pi\epsilon)}\eta^{-d+2}\Big(\frac{in}{\epsilon}\Big).
\end{equation}
The factor $R_\epsilon$ in front of this equation imposes a minimum value for the electric field such that for the field above this value the string pair creation is assumed to takes place. Indeed, one finds
\begin{equation}\label{43}
E_{min}=\pm T\sin\theta.
\end{equation}
The minus sign can be interpreted either as field with reversed direction or as a system with angle $-\theta$. Taking into account the maximum limit for the electric field, i.e. $E_{max}=T$ imposed by the Born-Infeld like factor in front of (38) we find the range for the electric field within which the pair creation occurs, as
\begin{equation}\label{44}
T\sin\theta<E<T.
\end{equation}
In the limit $\theta\rightarrow \frac{\pi}{2}$ the background field reaches its possible maximum value and as a result $\lim_{\theta\rightarrow \frac{\pi}{2}}R=0$ and $\lim_{\theta\rightarrow\frac{\pi}{2}} \epsilon=\frac{1}{2}$, which entails $w(\frac{\pi}{2})=0$. For the case $E<E_{min}$, i.e. $|\alpha|\leq 1$, we find the partition function upon substituting $\epsilon\rightarrow i\gamma$ in (44) as
\begin{equation}\label{39}
\mathcal Z=R_\gamma\sqrt\frac{T}{2s}\frac{q^{\frac{1}{2\pi}TY^2-\frac{d-2}{24}-\frac{1}{2}\gamma^2}}{\sinh(\frac{s\gamma}{2})}
\prod_{n=1}^\infty(1-q^{n})^{-d+4}(1-q^{n+\gamma})^{-1}(1-q^{n-\gamma})^{-1},
\end{equation}
with
\begin{equation}\label{1}
R_\gamma=\frac{1}{2\sin\theta}\sqrt{\sin^2\theta-\frac{E^2}{T^2}}.
\end{equation}
Contrary to the previous case ($\alpha<1$) there are no singularities along the integration contour. Therefore, the vacuum amplitude acquires no imaginary part which is meant as the absence of vacuum decay into string pairs. Again, the zero field limit, Eq.(25), is recovered via $R_\gamma\rightarrow\frac{1}{2}$ and $\gamma\rightarrow a$.
\section*{\large Conclusions}
As the case of point particle physics, string theory vacuum becomes instable in presence of an external electric field and decays into string pairs. We analyzed this problem for a system of angled D1-branes with bosonic string stretched between them and electric field along one of the D1-branes in frame work of the path integral formalism. We derived the string pair creation rate for this system. It seems that there is an angle dependent minimum value for the external field and pair creation occurs when the external field exceeds this minimum value. We also pointed out that the vacuum becomes stable and string pair creation vanishes as $\theta\rightarrow \frac{\pi}{2}$.
\section*{\large Appendix}
For the matrix $\mathcal{O}$ defined as $\mathcal{O}=\mathcal O_{1}+\mathcal O_{2}$ with
\begin{equation}\label{45}
\mathcal{O}_{1}=
\left(\begin{array}{ccc}
\textbf{A}&&\\
&\textbf{B}&\\
&&\textbf{C}
\end{array}\right),\qquad
\mathcal{O}_{2}=\left(\begin{array}{ccc}
0&-\textbf{I}&0\\
\textbf{I}&0&\textbf{J}\\
0&-\textbf{J}^{\scriptsize\textrm{t}}&0
\end{array}\right)
\end{equation}
we can write its determinant as
\begin{equation}\label{46}
\det\mathcal O=\det\mathcal{O}_{1}e^{\scriptsize{\textrm{Tr}}\ln(1+\mathcal{O}^{-1}_{1}\mathcal{O}_{2})}.
\end{equation}
The diagonal matrices in $\mathcal O_{1}$ have the generic form $\textbf{Q}=\textrm{diag}(q_0,q_1,\ldots)$. Now, from $\ln(1+x)=-\sum_{n=1}\frac{(-1)^n}{k}x^n$ and with the aid of
\begin{equation}\label{47}
\textrm{Tr}(\mathcal{O}^{-1}_{1}\mathcal{O}_{2})^{2n}=2(-1)^n\textrm{Tr}\Big[\textbf{B}^{-1}(\textbf{I}\textbf{A}^{-1}
\textbf{I}+\textbf{J}\textbf{C}^{-1}\textbf{J}^{\scriptsize\textrm{t}})\Big]^n,
\end{equation}
and by observing that for any diagonal matrix $\textbf{Q}$ we have
{\setlength\arraycolsep{2pt}
\begin{eqnarray}\label{48}
\textrm{Tr}(\textbf{Q}\textbf{I})^{n}&=&(c\textrm{Tr}\textbf{Q})^{n},\\
\textbf{I}\textbf{Q}\textbf{I}&=&(c\textrm{Tr}\textbf{Q})\textbf{I},\\
\textbf{J}\textbf{Q}\textbf{J}^{\scriptsize\textrm{t}}&=&(c\textrm{Tr}^\prime\textbf{Q})\textbf{I}.
\end{eqnarray}}
where we have defined $\textrm{Tr}\textbf{Q}=\sum_{n=0}q_n$ and $\textrm{Tr}^{\prime}\textbf{Q}=\sum_{n=1}q_n$ we get
\begin{equation}\label{49}
\textrm{Tr}\ln(1+\mathcal{O}^{-1}\mathcal{O})=\ln\Big[1+c^{\,2}\textrm{Tr}\textbf{B}^{-1}(\textrm{Tr}\textbf{A}^{-1}
+\textrm{Tr}^\prime\textbf{C}^{-1})\Big].
\end{equation}
Note that here we have denoted the factor $E_m$ in (33) with $c$. Therefore, we find the determinant as
\begin{equation}\label{50}
\det\mathcal O=\det\textbf{A}\det\textbf{B}\det\textbf{C}\Big[1+c^{\,2}\textrm{Tr}\textbf{B}^{-1}(\textrm{Tr}\textbf{A}^{-1}
+\textrm{Tr}^\prime\textbf{C}^{-1})\Big].
\end{equation}
In particular with $\textbf{A}=\textbf{M}^a_m$ and $\textbf{C}=\textbf{M}^{-a}_m$ one finds
\begin{equation}\label{51}
\textrm{Tr}(\textbf{M}^a_m)^{-1}+\textrm{Tr}^\prime(\textbf{M}^{-a}_m)^{-1}=\frac{1}{\frac{1}{4}{sT}}\sum_{n=-\infty}^{\infty}\frac{1}{\lambda^a_{mn}}.
\end{equation}
To evaluate the infinite sum of (60) one first writes
\begin{equation}\label{52}
\sum_{n=-\infty}^{\infty}\frac{1}{\lambda^a_{mn}}=\frac{\partial}{\partial\omega^2_m}\sum_{n=-\infty}^{\infty}\ln \lambda^a_{mn}=
\frac{\partial}{\partial\omega^2_m}\ln\prod_{n=-\infty}^{\infty}\lambda^a_{mn}.
\end{equation}
The infinite product can be calculated straightforwardly by invoking the formula
\begin{equation}\label{53}
\prod_{m\in\mathbb Z}(mx+y)=2\sinh\bigg(\frac{i\pi y}{x}\bigg).
\end{equation}
The final result is
\begin{equation}\label{54}
\sum_{n=-\infty}^{\infty}\frac{1}{\lambda^a_{mn}}=\frac{\pi}{2\omega_m}\big[\coth\pi(\omega_m+ia)+\coth\pi(\omega_m-ia)\big].
\end{equation}
\section*{\large References}
[1]\hspace{0.2cm}E. S. Fradkin, A. A. Tseytlin, Phys. Lett. \textbf{B163} (1985) 123.\\\
[2]\hspace{0.2cm}A. Abouelsaood, C. G. Callan, C. R. Nappi and S. A. Yost, Nucl. Phys.
\textbf{B280} (1987) 599.\\\
[3]\hspace{0.2cm}V. V. Nesterenko, Int. J. Mod. Phys. \textbf{A4} (1989) 2627.\\\
[4]\hspace{0.2cm}M. B. Green, J. H. Schwarz and E.Witten, \textit{Superstring Theory}, Cambridge University
Press, 1987.\\\
[5]\hspace{0.2cm}C. P. Burgess, Nucl. Phys. \textbf{B294} (1987) 427.\\\
[6]\hspace{0.2cm}C. Bachas, M. Porrati, Phys. Lett. \textbf{B296} (1992) 77.\\\
[7]\hspace{0.2cm}C. Bachas, Phys. Lett. \textbf{B374} (1996) 37.\\\
[8]\hspace{0.2cm}A. A. Bytsenko, S.D. Odintsov and L. Granda, Mod. Phys. Lett. \textbf{A11} (1996) 2525.\\\
[9]\hspace{0.2cm}M. Porrati, arXiv:hep-th/9309114. \\\
[10]\hspace{0.2cm}C. Acatrinei, Nucl. Phys. \textbf{B539} (1999) 513. \\\
[11]\hspace{0.2cm}C. Acatrinei, R. Iengo, Phys. Lett. \textbf{B482} (2000) 420.\\\
[12]\hspace{0.2cm}J. Ambjorn, Y. M. Makeenko, G. W. Semenoff and R. J. Szabo, JHEP \textbf{0302} (2003) 026. \\\
[13]\hspace{0.2cm}J. Schwinger, Phys. Rev \textbf{D82} (1951) 664.\\\
[14]\hspace{0.2cm}J. H. Cho, P. Oh, C. Park and J. Shin,
JHEP \textbf{0505}, (2005) 004.\\\
[15]\hspace{0.2cm}J. X. Lu, Shan-Shan Xu, JHEP \textbf{0909} (2009) 093. \\\
[16]\hspace{0.2cm}A. Jahan, Mod. Phys. Lett. \textbf{A25} (2010) 619.\\\
[17]\hspace{0.2cm}A. Matusis, Int. J. Mod. Phys. \textbf{A14} (1999) 1153.\\\
[18]\hspace{0.2cm}H. Arfaei, M. M. Sheikh-Jabbari, Phys. Lett. \textbf{B394} (1997) 288.\\\
[19]\hspace{0.2cm}T. Kiato, N. Ohta and J. Zhaou, Phys. Lett. \textbf{B428} (1998) 68.\\\
[20]\hspace{0.2cm}D. Bak, N. Ohta, Phys. Lett. \textbf {B527} (2002) 131.\\\
[21]\hspace{0.2cm}J. H. Cho and P. Oh, JHEP \textbf {0301} (2003) 046.\\\
[22]\hspace{0.2cm}R. C. Myers, D. J.Winters, JHEP \textbf {0212} (2002) 061.
\end{document}